\documentclass[a4paper,11pt]{article}
\pdfoutput=1 

\usepackage{jheppub} 
\usepackage{mathtools}

\usepackage[T1]{fontenc} 

\title{Impurities in Schrödinger field theories\\and s-wave resonance}

\author[1]{Avia Raviv-Moshe}
\author[1,2]{Siwei Zhong}

\affiliation[1]{Simons Center for Geometry and Physics, SUNY,\\Stony Brook, NY 11794, USA}
\affiliation[2]{C. N. Yang Institute for Theoretical Physics, Stony Brook University,\\Stony Brook, NY 11794, USA}

\emailAdd{araviv-moshe@scgp.stonybrook.edu}
\emailAdd{siwei.zhong@stonybrook.edu}

\abstract{We study point impurities in non-relativistic quantum field theories, with a focus on scale-invariant fixed points. We establish the framework of conformal defects in Schrödinger field theories and their correspondence to many-body states in a harmonic trap. We discuss a multi-critical quantization of the one-body wave function induced by an impurity, with examples of free and unitary Fermi gases. Our results can potentially be tested in experimental setups. }

\begin{document} 
\maketitle
\flushbottom

\section{Introduction}

Defects provide important new perspectives in modern understanding of quantum field theories. In recent years, defects in relativistic Conformal Field Theories (CFTs) have been extensively studied due to their immense relevance to physical systems in nature (see e.g. \cite{gaiotto2014bootstrapping,parisen2022boundary,cuomo2022localized,cuomo2022renormalization,cuomo2022spin,giombi2023line, aharony2023phases}). In this letter, we study defects in non-relativistic CFTs. Such a setup has been rarely explored in the quantum field theory framework \cite{gupta2022non}, yet it possesses vast experimental applications (see, for example, \cite{ cucchietti2006strong, PhysRevLett.102.230402, modugno2010anderson, zipkes2010trapped, PhysRevA.83.033619, massignan2021universal,targonska2010self,jiang2011single, massignan2014polarons}).
We focus on non-relativistic CFTs that capture the long-distance limit of many-body wave functions at scale-invariant fixed points, where Schrödinger symmetry is preserved \cite{nikolic2007renormalization,nishida2007fermi,kaplan2009conformality,nishida2011unitary}.   

The Schr\"{o}dinger group $Sch(d)$ of the $(d+1)$-dimensional spacetime $(t,x)$ is spanned by transformations that read \cite{Niederer:1972zz, PhysRevD.5.377, Henkel:2003pu, pal2018unitarity}
\begin{equation} 
\label{eq_Schrod_general_trans_rule}
    t \to \frac{at+b}{ct+d}\,, \qquad x\to \frac{U\cdot x+v t+\rho}{ct+d}~,
\end{equation}
where $a,b,c,d \in \mathbb{R}$ and $ad-bc=1$ generates the $SL(2,\mathbb{R})$ subgroup. In \eqref{eq_Schrod_general_trans_rule}, $U\in O(d)$ is a spatial rotation/reflection matrix, $v\in \mathbb{R}^d$ marks the velocity of the Galilean boosts, and $\rho \in \mathbb{R}^d$ represents spatial translations. $Sch(d)$ derives its name from being the symmetry group of the free particle Schr\"{o}dinger equation in $d$-dimensional space. Certain field theories are known to demonstrate the $Sch(d)$ invariance, including a free field realization. In this paper, we focus on such realizations and study the free field theory as well as the dilute Fermi gas at infinite scattering length \cite{nikolic2007renormalization,nishida2007fermi,kaplan2009conformality,nishida2011unitary}, which is an interacting Schrödinger field theory with various applications.

In the context of ultra-cold atomic gases, the interacting theories we study are also commonly referred to as unitary Fermi gases. Such models describe neutral particles with short-range attractive interactions near their Feshbach s-wave resonance. In experimental setups, the s-wave scattering length $a$ of spin-$\frac{1}{2}$ fermionic atoms in $d=3$ space can be tuned through an external magnetic field \cite{regal2004observation,zwierlein2004condensation}. When $|a|\gg k_\text{F}^{-1}$, such that it exceeds the inter-particle distance of the Fermi momentum scale $k_\text{F}$, the system approaches the unitary limit where the $Sch(3)$ conformal symmetry emerges. This limit can also be understood as the strongly coupled crossover between two perturbative regimes, namely the Bardeen–Cooper–Schrieffer pair condensate ($a<0$ and $|ak_\text{F}|\ll 1$) and the Bose-Einstein condensate ($a>0$ and $|ak_\text{F}|\ll 1$) \cite{nikolic2007renormalization,nishida2011unitary}. Another important application of non-relativistic field theories is in nuclear scattering processes. Particularly, unitary Fermi gas models are used to describe neutron-neutron scatterings in $d=3$ \cite{kaplan1998new, bedaque1999renormalization,hammer2021unnuclear,hongo2022universal}. The s-wave scattering length of neutron $a_{\text{nn}}\approx -18.5 \text{fm}$ is found to be significantly larger than the scale associated with nuclear interactions $r_{\text{nn}}\approx 2.8\text{fm}$ \cite{vslaus1989neutron}. For wavelength $k$ in the window $r_{\text{nn}} \ll k^{-1}\ll |a_{\text{nn}}|$, the infinite scattering length limit serves as a useful approximation for effective field theory to make quantitative predictions. In this limit, observables are strongly constrained by the conformal invariance of $Sch(3)$, and universality classes emerge. 

Jackiw-Pi models in $d=3$ \cite{jackiw1990soliton,jackiw1990classical} mark another notable example of Schrödinger field theory. These models describe the kinematics of abelian \cite{doroud2016superconformal} and non-abelian \cite{doroud2018conformal} anyons at the low-energy limit, whose spectrum is subject to the $Sch(3)$ symmetry. We also remark that the $Sch(d)$ group is isomorphic to the symmetry group of a $d$-dimensional harmonic trap. Many-body wave functions in the trap belong to the same universality classes as field theories \cite{nishida2007nonrelativistic,werner2006unitary,werner2006unitaryap}, which led to a considerable experimental interest (see e.g. \cite{busch1998two,gharashi2015one}). 

This paper is devoted to spatial point impurities in Schrödinger field theories. Impurities are subjects with a renowned history of research (see, for example, \cite{de1934electrical,kondo1964resistance,sachdev1999quantum,vojta2000quantum} and references therein). In atomic gas systems, examples of point impurities include neutral atoms immersed in Bose-Einstein condensates \cite{kalas2006interaction,cucchietti2006strong,bruderer2008self,tempere2009feynman} and in Fermi superfluids \cite{targonska2010self,jiang2011single,vernier2011bound, massignan2014polarons}. Another way of creating impurities is by localized external fields, such as laser beams \cite{madison2000vortex}. Microscopic details of the impurities and their ambient systems vary in these examples, and it takes a case-by-case analysis when observables are concerned. However, certain universalities can be obtained for impurities embedded in Schrödinger field theories, as we elaborate in 
this paper. 

From a spacetime point of view, spatial point impurities are line defects extended in the time direction. Even when the ambient bulk theory is at a scale-invariant
fixed point, defect parameters are generically renormalized under the coarse-graining procedure known as the defect Renormalization Group (RG) flow (see e.g. \cite{cuomo2022renormalization} and references therein). Fixed points of the defect RG flow preserve $SL(2,\mathbb{R})\subseteq Sch(d)$, which imposes strong constraints on the impurity spectrum. Remarkably, $SL(2,\mathbb{R})$ invariance enables us to characterize the defect universality class with relatively few conformal data, i.e., the conformal dimensions and OPE coefficients of operators. We also note that, for simplicity, only isotropic impurities that preserve $O(d)\subseteq Sch(d)$ are considered in this paper. Our analysis also applies to cases where the microscopic impurity is 
 spatially anisotropic yet $O(d)$ emerges in the infrared. 

The rest of the paper is organized as follows. First, we develop the Defect Conformal Field Theory (DCFT) formalism for generic impurities at RG fixed points, where the bulk-to-defect OPE and the state/operator correspondence are discussed. With mild assumptions on locality, we prove the existence of a class of fixed points that corresponds to the Feshbach s-wave resonance of the impurity and obtain analytical results concerning the one-body wave function. We then study the models of free and unitary Fermi gases, where the defect is induced by an impurity contact interaction. We show that in Fermi gas models, indeed a fine-tuning of the impurity parameter leads to the s-wave resonance fixed point. We conclude this paper with the impurity phase diagram and future directions.

\section{DCFT Generalities}
\label{sec_DCFT Generalities}

\subsection{Review of Schrödinger Field Theories}
We start by reviewing the properties of the Schrödinger field theory in the ambient bulk. Simply connected parts of $Sch(d)$ in \eqref{eq_Schrod_general_trans_rule} are generated by $\mathbb{R}^d$ spatial translations $P_\mu$, $SO(d)$ rotations $J_{\mu \nu}$, $\mathbb{R}^d$ Galilean boosts $K_\mu$, time translation $H$, special conformal transformation $C$, and $z=2$ anisotropic dilation $D$\footnote{ $z$ is commonly referred as the dynamical critical exponent, such that upon a scaling transformation $t\to \lambda^{z}t$ and $x \to \lambda x$.}, where we have denoted spatial indices $\mu, \nu =1,2,...,d$. The central extension $M\in \mathbb{R}$ between translations and boosts are of particular interest:
\begin{equation}
\label{eq_central extension def}
   \left[ K_\mu, P_\nu\right]=i\delta_{\mu\nu}M~.
\end{equation}
$M$ in \eqref{eq_central extension def} is identified as the non-relativistic mass, generating the $U(1)_M$ particle number symmetry. In open systems, particle pumps or sinks violate the $U(1)_M$ symmetry. We emphasize that all examples considered in this paper are closed, static, and preserve $U(1)_M$.

Operators are categorized by their representation under $Sch(d)$, which also governs their dynamics. The particle number $m_{\mathcal{O}}$ of an operator $\mathcal{O}$ is identified as its charge under $U(1)_M$, such that $[M,\mathcal{O}]=-im_{\mathcal{O}} \mathcal{O}$. On the other hand, an operator $\mathcal{O}(t,x)$ with scaling dimension (critical exponent) $\Delta(\mathcal{O})$ satisfies:
\begin{equation}
\label{eq_dilation def}
\left[D, \mathcal{O}\right] = i(2t\partial_t+x\cdot\partial+\Delta(\mathcal{O}))\mathcal{O}~.
\end{equation}
When acting on $\mathcal{O}$, $H$ and $P_\mu$ raise its dimension by $2$ and $1$, while $C$ and $K_\mu$ reduce it by $2$ and $1$, respectively. Unitarity prohibits negative norm states in the Hilbert space and implies a lower bound on non-identity operator dimensions $\Delta(\mathcal{O})\geq \frac{d}{2}$ \cite{Nishida:2010tm}. Primary operators are identified as the lowest-lying operators in a given representation of the Schrödinger group, such that they satisfy $\left[ K_\mu, \mathcal{O} \right] = \left[ C, \mathcal{O} \right] =0$ \cite{nishida2007nonrelativistic} \footnote{The notion of primary operators suffers from ambiguity when $m_{\mathcal{O}}=0$, where $P_\mu$ and $K_\mu$ commute. To clarify, we exclusively use the term bulk/defect primaries when their charge under particle number symmetry is non-zero.}, where $\left[K_\mu, \mathcal{O} \right]= \left(-it\partial_\mu+x_\mu m_{\mathcal{O}}\right)\mathcal{O}$ and 
\begin{equation}
\label{eq_special conformal transformation}
\left[C, \mathcal{O}\right] =\left(-itx \cdot \partial-it^2\partial_t-it\Delta (\mathcal{O})+\frac{|x|^2}{2} m_{\mathcal{O}} \right)\mathcal{O}.
\end{equation}
In what follows, we will often use Euclidean time formalism by Wick rotating $\tau=it$. This provides us with calculation convenience as the convergence of certain functions becomes readily apparent in the Euclidean time. 

In the non-relativistic regime, anti-particles decouple from the spectrum. Consequently, normal ordering for composite operators can be done without encountering singularities \cite{bergman1992nonrelativistic}. 
For a non-identity operator $\mathcal{O}\neq \mathbb{I}$, either the operator annihilates the vacua $\mathcal{O}|0\rangle =0$ or its conjugate does $\mathcal{O}^\dagger |0\rangle =0$.\footnote{This is an assumption that holds true in all examples of this paper. In general, there could be $U(1)_M$ neutral field contents (e.g. the GED \cite{festuccia2016symmetries,chapman2020renormalization}). Nevertheless, matter field propagators in GED satisfy the Euclidean time
ordering.} Let us consider a scalar creation(annihilation) operator $\mathcal{O}^\dagger$($\mathcal{O}$). The two-point function $\langle \mathcal{O}^\dagger(\tau_1)\mathcal{O}(\tau_2)\rangle \propto \Theta(\tau_{12})$ yields Euclidean time ordering \cite{pal2018unitarity}, where $\Theta(\tau)$ is the Heaviside step function. Furthermore, $\langle \mathcal{O}^\dagger\mathcal{O}\rangle$ is constrained by \eqref{eq_dilation def}, \eqref{eq_special conformal transformation}, translational and rotational symmetry of $Sch(d)$, and is fixed up to an overall coefficient \cite{nishida2007nonrelativistic}. An orthogonal basis of the bulk scalar primary operators can be chosen, such that
 \begin{equation}
\label{eq_bulk primary basis}
    \langle \mathcal{O}_1 (\tau_1,x_1)\mathcal{O}_2^\dagger(\tau_2,x_2)  \rangle=\delta_{\mathcal{O}_1,\mathcal{O}_2}\frac{\Theta(\tau_{12})}{\tau_{12}^{\Delta(\mathcal{O})}}e^{-\frac{m_{\mathcal{O}}|x_{12}|^2}{2\tau_{12}}},
\end{equation}
where $\tau_{ij}\equiv\tau_i-\tau_j$ and $x_{ij}\equiv x_i-x_j$. Three-point and higher-point correlations depend on $Sch(d)$ crossratios and their functional forms can not be uniquely determined from the Schr\"{o}dinger symmetry.

Another key aspect of Schr\"{o}dinger field theories is the state/operator correspondence \cite{nishida2007nonrelativistic,goldberger2015ope}. We introduce the oscillator frame $(\Tilde{t}, \Tilde{x}_\mu)$ with the harmonic frequency $\omega$:
\begin{equation}
\label{eq_oscillator frame}
    \Tilde{t}=\frac{1}{\omega}\arctan{(\omega t)}~, \quad 
    \Tilde{x}_\mu=\frac{x_\mu}{\sqrt{1+\omega^2 t^2}}.
\end{equation}
whose time translation is generated by $\Tilde{H}=H+\omega^2 C$. $\Tilde{H}$ marks the Hamiltonian of particles in a harmonic trap. An operator $\mathcal{O}^\dagger$ under the correspondence is mapped to an energy eigenstate $|\mathcal{O}^\dagger\rangle$ as follows
\begin{equation}
\label{eq_state/operator correspondence}
\begin{aligned}
        &|\mathcal{O}^\dagger\rangle \equiv e^{-H/\omega}\mathcal{O}^\dagger(t=0,x=0)\left|0\right\rangle~,\\
        &\text{such that }\Tilde{H} |\mathcal{O}^\dagger\rangle =\omega \Delta(\mathcal{O})|\mathcal{O}^\dagger\rangle~.
    \end{aligned}
\end{equation}
The scaling dimension \eqref{eq_dilation def} in is interpreted as the energy level in units of the harmonic frequency $\omega$. On the other hand, the two-point correlation function is mapped to the wave function $\Psi_\mathcal{O}(\Tilde{\tau}, \Tilde{x})\equiv \langle 0|\mathcal{O}(\Tilde{\tau}, \Tilde{x}) |\mathcal{O}^\dagger\rangle$. Note that the Heisenberg picture operator $\mathcal{O}^\dagger$ in \eqref{eq_state/operator correspondence} is inserted at the Euclidean time $\tau=-1/\omega$ of the Galilean frame, which in the oscillator frame is the infinite past $\Tilde{\tau}=-\infty$. The two-point function in the oscillator frame reads \cite{goldberger2015ope}
\begin{equation}
\Psi_\mathcal{O}(\Tilde{\tau}, \Tilde{x}) \propto e^{-\omega \Delta(\mathcal{O})\Tilde{\tau}-\frac{m_{\mathcal{O}}}{2}\omega|\Tilde{x}|^2}~,
\end{equation}
 which is identified as the Hartree-Fock wave function of the many-body bound state $|\mathcal{O}^\dagger\rangle$ in the harmonic trap.

\subsection{Bulk-to-Defect OPE}
We now develop the formalism of studying line defects (impurities) in a Schrödinger field theory. Particularly, we focus on the defects that preserve the residual conformal group $SL(2,\mathbb{R})$, the transversal rotational group $O(d)$, and the particle number symmetry $U(1)_M$. Without losing generality, we assume the defect is located at $x=0\in \mathbb{R}^d$ in space. We use the hat superscript to denote a defect operator $\hat{O}_l$, where $l\in \mathbb{N}$ marks its $SO(d)$ spin. Similar to \eqref{eq_bulk primary basis}, an orthogonal basis can also be chosen among defect $SL(2,\mathbb{R})$ primary operators, such that 
\begin{equation}
\label{eq_defect primary basis}
    \langle \hat{\mathcal{O}}_{1,l_1,q_1} (\tau_1)\hat{\mathcal{O}}^\dagger_{2,l_2,q_2} (\tau_2)  \rangle=\delta_{\hat{\mathcal{O}}_{1},\hat{\mathcal{O}}_{2}}\delta_{q_1,q_2}\frac{\Theta(\tau_{12})}{\tau_{12}^{\Delta(\hat{\mathcal{O}}_{l})}}~,
\end{equation}
where $q_1,q_2$ denote the $SO(d)$ spin indices. We remark that the defect unitarity bound for (non-identity) operators follows from $SL(2,\mathbb{R})$, and it reads $\Delta(\hat{\mathcal{O}}_l)> 0$.

The two-point function between a bulk scalar operator $\mathcal{O}$ and a defect operator $\hat{O}_{l}$ is subject to the $SL(2,\mathbb{R})\times SO(d)$ invariance. In particular, \eqref{eq_dilation def} and \eqref{eq_special conformal transformation} can be used to fix $\langle \mathcal{O} \hat{\mathcal{O}}^\dagger_{l,q} \rangle$ up to an overall constant (ie, the OPE coefficient) $C^{\mathcal{O}}{}_{\hat{\mathcal{O}}_{l}}\in \mathbb{C}$, and $U(1)_M$ further requires that $C^{\mathcal{O}}{}_{\hat{\mathcal{O}}_{l}}\propto \delta_{m_{\mathcal{O}},m_{\hat{\mathcal{O}}_{l}}}$.\footnote{
We argue that bulk one-point functions vanish identically, even with the defect's presence. This fact is evident to operators charged under $U(1)_M$. For neutral operators, a perturbative proof follows from normal ordering and the absence of anti-particles in the spectrum. We note that a complete non-perturbative proof of this statement remains open.} Let $r\in \mathbb{R}^+$ be the spatial distance to the impurity, 
$\Omega\in \mathbb{S}^{d-1}$ be the angular coordinate and $Y_{l,q}(\Omega)$ be the spherical harmonics \eqref{appendix_spherical harmonics}, we find
\begin{equation}
\label{eq_bulk-defect two point function}
    \langle \mathcal{O} (\tau_1 ,r,\Omega ) \hat{\mathcal{O}}^\dagger_{l,q} (\tau_2)  \rangle=C^{\mathcal{O}}{}_{\hat{\mathcal{O}}_{l}}\frac{\Theta(\tau_{12})Y_{l,-q}(\Omega)e^{-\frac{m_{\mathcal{O}}r^2}{2\tau_{12}}}}{\tau_{12}^{\Delta(\hat{\mathcal{O}}_{l})}r^{\Delta(\mathcal{O})-\Delta(\hat{\mathcal{O}}_{l})}}~.
\end{equation}
With a fixed normalization in \eqref{eq_bulk primary basis} and \eqref{eq_defect primary basis}, the OPE coefficients $C^{\mathcal{O}}{}_{\hat{\mathcal{O}}_{l}}$ encode physical observables of the impurity. When a bulk operator $\mathcal{O}^\dagger$ approaches the defect at $r=0$, from equations \eqref{eq_defect primary basis} and \eqref{eq_bulk-defect two point function} we obtain the OPE decomposition:
\begin{equation}
\begin{aligned}
    \label{eq_bulk-to-defect OPE}
&\mathcal{O}^\dagger (\tau ,r,\Omega )=\sum_{\hat{\mathcal{O}}_{l,q}}\frac{C^{\mathcal{O}}{}_{\hat{\mathcal{O}}_{l}}Y^*_{l,q}(\Omega)}{r^{\Delta(\mathcal{O})-\Delta(\hat{\mathcal{O}}_{l})}} \mathcal{B}(r,\partial_\tau)\hat{\mathcal{O}}^\dagger_{l,q}(\tau)~,\\
&\text{where }\mathcal{B}_\mathcal{O}=\sum_{n\in \mathbb{N}}\frac{(-m_\mathcal{O}/2)^n}{n!(\Delta(\hat{\mathcal{O}}_l))_n} r^{2n}\partial^n_\tau~,
\end{aligned}
\end{equation}
and $(\Delta)_n$ is the Pochhammer symbol. In higher-point correlation functions, we leave it as a conjecture that there exists a defect limit where \eqref{eq_bulk-to-defect OPE} is valid as an operator equation. The convergence condition of \eqref{eq_bulk-to-defect OPE}, as well as its associated crossing symmetry equation (tantamount to the relativistic case \cite{liendo2013bootstrap}), goes beyond the scope of this paper. We refer to \cite{goldberger2015ope} for a relevant study.

%\textcolor{red}{For higher-point correlation functions, we assume the existence of a defect limit where the bulk operator is infinitely close to the defect and the decomposition \eqref{eq_bulk-to-defect OPE} is valid.} 
%\textcolor{red}{The OPE decomposition \eqref{eq_bulk-to-defect OPE} is valid at the two-point level, and is only conjectural beyond this order. Whether \eqref{eq_bulk-to-defect OPE} truly holds as an operator equation is an open question that goes beyond the scope of this paper. For analyses and calculations performed in this paper, however, \eqref{eq_bulk-to-defect OPE} is used only at the two-point level, hence, such an assumption will not be required. }

\subsection{State/Operator Correspondence}

Defect operators $\hat{\mathcal{O}}_l$ under the correspondence are mapped to energy eigenstates in a harmonic trap deformed by delta-function potentials localized at the center. In physical scenarios, such a localized potential can be induced by heavy neutral atoms \cite{cucchietti2006strong} located at the trap center. 

To demonstrate the correspondence, we first note that the defect preserves $SL(2,\mathbb{R})\subseteq Sch(d)$. We can therefore construct an eigenstate $|\hat{\mathcal{O}}^\dagger_l\rangle=e^{-H/\omega} \hat{\mathcal{O}}^\dagger_l(t=0)|0\rangle$, similar to \eqref{eq_state/operator correspondence}, such that in the oscillator frame \eqref{eq_oscillator frame} $\Tilde{H}|\hat{\mathcal{O}}^\dagger_l\rangle=\omega \Delta(\Tilde{\mathcal{O}_l})|\hat{\mathcal{O}}^\dagger_l\rangle$. Hence, the scaling dimensions of defect operators bear a natural interpretation as energy levels in units of harmonic frequency $\omega$. Furthermore, let us define the wave function $\Psi_{\hat{\mathcal{O}}_{l,q}}\equiv\langle 0 |\mathcal{O}(\Tilde{\tau},\Tilde{r},\Omega)|\hat{\mathcal{O}}^\dagger_l\rangle$. From the two-point function \eqref{eq_bulk-defect two point function}, we find:
\begin{equation}\label{eq_defect wave function}
\Psi_{\hat{\mathcal{O}}_{l,q}}(\Tilde{\tau}, \Tilde{r},\Omega) \propto e^{-\omega \Delta(\hat{\mathcal{O}}_{l})\Tilde{\tau}}\frac{Y^*_{l,q}(\Omega) e^{-\frac{m_{\mathcal{O}}}{2}\omega\Tilde{r}^2}}{\Tilde{r}^{\Delta(\mathcal{O})-\Delta(\hat{\mathcal{O}}_{l})}} ~,
\end{equation}
which is indeed the Hartree-Fock wave function of spin-$l$ and has an asymptotical central fall-off of $\Tilde{r}^{\Delta(\hat{\mathcal{O}}_{l})-\Delta(\mathcal{O})}$. In the Galilean frame, defect descendants $\partial_\tau^{n} \hat{\mathcal{O}}_l$ are generated by repeatedly acting with $H$ on the primary $\hat{\mathcal{O}}_l$. We remark that in the oscillator frame, such descendants are mapped to excitations of the breathing modes \cite{werner2006unitary,werner2006unitaryap}, created($+$) and annihilated($-$) by the ladder operator $\Tilde{L}_\pm\equiv \frac{H}{2\omega}-\frac{\omega}{2}C\pm iD$.

\subsection{One-Body Sector}

Let us discuss general analytical properties concerning the defect operators of $U(1)_M$ unit charge \footnote{When there are multiple particle species with masses $m_1, m_2,...$, we refer to irreducible operators with corresponding $U(1)_M$ charges $m_1, m_2,...$}. We focus on systems in spatial dimension $d\neq 2$ described by a generic local action. Our following analysis is general and it applies in a variety of different models, including the Fermi gases \eqref{eq_general fermi lagragian}.

For concreteness, we present our argument with reference to a bosonic theory of $\phi \in \mathbb{C}$ in $d$ dimension:
\begin{equation}
\label{eq_example largragian}
\mathcal{L}=\phi^\dagger\partial_\tau \phi+\frac{|\nabla \phi|^2}{2m}+\mathcal{L}_\text{int}(\phi)+\delta^d(x)\mathcal{L}_{\text{defect}}(\phi)~,
\end{equation}
where both the bulk ($\mathcal{L}_\text{int}$) and the defect ($\mathcal{L}_\text{defect}$) local interactions perserve $U(1)_M$ and $SL(2,\mathbb{R})$. The bulk one-body operator has a protected dimension $\Delta(\phi)=\frac{d}{2}$, regardless of the interaction $\mathcal{L}_\text{int}$ \cite{nishida2007nonrelativistic, Nishida:2010tm}. This fact follows the equation of motion $(\partial_{\tau}-\frac{\nabla^2}{2m})\phi=-\partial_{\phi^\dagger} \mathcal{L}_\text{int}\propto \phi^\dagger(...)$ and normal ordering, such that $\langle 0 |(\partial_{\tau}-\frac{\nabla^2}{2m})\phi=0$. With the bulk-to-defect two-point function \eqref{eq_bulk-defect two point function}, we obtain
\begin{equation}
\label{eq_bulk-to-defect EoM}
\left(\partial_{\tau_1}-\frac{\nabla^2}{2m}\right)\langle \phi(\tau_1, r, \Omega) \hat{\phi}_{l,q}^\dagger(\tau_2)\rangle
\propto \left(\Delta (\hat{\phi}_l)+\frac{d}{2}+l-2\right)\left(\Delta (\hat{\phi}_l)-\frac{d}{2}-l\right)=0~.
\end{equation}
Therefore, we conclude that defect primaries in the OPE of one body operators are of dimension either $\Delta(\hat{\phi}_l)=\frac{d}{2}+l$ or $\Delta(\hat{\phi}_l)=2-\frac{d}{2}-l$. This resembles the case studied in \cite{lauria2021line}, 
however, our statement also holds for interacting theories. While our analysis is general, two exceptional cases are $d=2$ and non-local actions. When $d=2$, fluxes of global symmetries can be attached to the impurity, and the Aharonov-Bohm effect leads to non-integer $l$ \cite{gaiotto2014bootstrapping, 
doroud2016superconformal,doroud2018conformal,bianchi2021monodromy}. A non-local $Sch(d)$ potential \cite{frank1971singular,beane2001singular,kaplan2009conformality,landau2013quantum} modifies the equation of motion of $\phi$ on the LHS of \eqref{eq_bulk-to-defect EoM}, and also leads to different defect primary dimensions. In both exceptional cases, the generalization of our analysis is straightforward but goes beyond the scope of this work.

Adopting the name convention from the AdS/CFT literature \cite{burges1986supersymmetry, DHoker:2002nbb}, we associate the $\Delta(\hat{\phi}_l)=\frac{d}{2}+l$ defect fixed points with the standard quantization schemes of the Schrödinger fields, and $\Delta(\hat{\phi}_l)=2-\frac{d}{2}-l$ fixed points with the alternative schemes. It is important to note that for a given defect theory, there exists one decisive quantization scheme for each bulk field, while there can be multiple defect primaries that satisfy either solutions of \eqref{eq_bulk-to-defect EoM}. For simplicity, we will take the assumption that the impurity has no internal Hilbert space, and the defect one-body primaries are determined by the quantization of the bulk wave functions. For $l\geq 2$, the alternative fixed point violates unitarity in any physical dimension $d$. A unitary alternative fixed point of the s-wave ($l=0$) resonance exists when $d<4$, and for the p-wave ($l=1$) resonance, it exists when $d<2$. Standard fixed points and alternative ones are connected by 1-loop exact renormalization flows, which are triggered by defect bilinear deformations \cite{gubser2003universal, nagar2024planar}. This will be further discussed in the following examples.

In this paper, we focus on s-wave resonance in spatial dimensions $2<d<4$, where the $d=3$ case is relevant to many physical setups of experimental importance. We note that unitarity requires standard quantization for $l \geq 1$ spin components, while for $l=0$ both quantizations are allowed. Let us consider the bulk-to-bulk two-point function $\langle \phi(\tau_1,x_1)\phi^\dagger(\tau_2,x_2)\rangle$. The one-body operators $\phi$ and $\phi^\dagger$ are subject to the equation of motion, and their correlation with the defect presence can be explicitly obtained. Note that the asymptotic condition for the correlation function is determined by the DCFT data as in \eqref{eq_bulk-defect two point function} and \eqref{eq_bulk-to-defect OPE} \footnote{
In Fourier space, the radial mode profile for the standard($+$) and alternative($-$) quantization reads $r^{1-\frac{d}{2}} J_{\pm(l+\frac{d}{2}-1)}(kr)$, where $k\in \mathbb{R}^+$ and $J_\nu$ are Bessel functions.}. For the fixed point $\Delta(\hat{\phi}_0)=\frac{d}{2}$, we obtain $\langle \phi(\tau_1,x_1)\phi^\dagger(\tau_2,x_2)\rangle=G_m(\tau_{12},x_{12})$, where
\begin{equation}
\label{eq_free field propagator}
G_{m}(\tau,x)=\Theta(\tau)\left(\frac{m}{2\pi \tau}\right)^{d/2}e^{-\frac{m x^2}{2\tau}}
\end{equation}
coincides with a free field propagator as in the case without an impurity \footnote{Equation \eqref{eq_free field propagator} indicates the particle-impurity two-body interaction resides at a trivial fixed point. This does not directly imply defect triviality, as imprints can be left in the many-body sector.}. On the other hand, for the s-wave resonance fixed point $\Delta(\hat{\phi}_0)=2-\frac{d}{2}$, we find
$\langle \phi(\tau_1,x_1)\phi^\dagger(\tau_2,x_2)\rangle=G_m(\tau_{12},x_{12})+D_m(\tau_{12},r_1,r_2)$. The ``defect contribution'' reads:
\begin{equation}
\label{eq_exact defect propagator}
D_m(\tau,r_1,r_2)=-\Theta(\tau)\pi ^{-\frac{d}{2}} \sin \left(\pi d /2 \right) \Gamma \left(d/2\right) \frac{m e^{-\frac{m
   }{2\tau}\left(r_1^2+r_2^2\right)} }{\pi \tau (r_1 r_2)^{\frac{d}{2}-1} }K_{1-\frac{d}{2}}\left(\frac{m r_1 r_2}{\tau }\right)~,
\end{equation}
where $K_\nu$ are modified Bessel functions of the second kind. Particularly, the correlation function of $d=3$ impurity s-wave resonance is such that:
\begin{equation}
\label{eq_exact defect propagator}
\left. \langle \phi(\tau_1,x_1)\phi^\dagger(\tau_2,x_2)\rangle\right|_{d=3}=\Theta(\tau_{12})\left(\frac{m}{2\pi\tau_{12}}\right)^{3/2}\left(e^{-\frac{m x_{12}^2}{2 \tau_{12} }}+\frac{\tau_{12}}{m r_1r_2}e^{-\frac{m (r_1+r_2)^2}{2 \tau_{12} }} \right)~.
\end{equation}
We emphasize that the above analysis, and particularly the two-point functions \eqref{eq_free field propagator} and \eqref{eq_exact defect propagator}, hold for any local interacting bulk theory. This includes the unitary Fermi gas that will be discussed in what comes next.

\section{Free Fermi Gas}
\label{sec_free fermi gas}

We move on to study the spatial point impurity in the $(d+1)$-dimensional Fermi gas, which is a model of wide experimental applications \cite{hammer2021unnuclear,hongo2022universal, busch1998two,zwierlein2004condensation}. The system contains two species of complex Grassmannian fields $\psi_\sigma$, which physically represent a spin-$\frac{1}{2}$ fermion with $\sigma=\uparrow,\downarrow$ in the fundamental representation of $Sp(2)\simeq SU(2)$. The model is associated with Lagrangian density \cite{nikolic2007renormalization,nishida2007fermi,nishida2007nonrelativistic,nishida2011unitary}:
\begin{equation}
\label{eq_general fermi lagragian}
\mathcal{L}=\sum_{\sigma=\uparrow,\downarrow}\psi^\dagger_{\sigma}\left(\partial_\tau-\frac{\nabla^2}{2m}+\mu_\text{b} \right)\psi_\sigma+\lambda_\text{b}\psi_\uparrow^\dagger\psi_\downarrow^\dagger\psi_\downarrow\psi_\uparrow~,
\end{equation}
where $\mu_\text{b}$ is the chemical potential and $\lambda_\text{b}$ marks the short-range interaction between fermions. For simplicity, we also denote the composite two-body operator $\Phi^\dagger \propto \psi_\uparrow^\dagger\psi_\downarrow^\dagger$ in what follows.

In the bulk, we fine-tune the chemical potential $\mu_\text{b}=0$, such that the theory is at the zero-density critical point. Let $\Lambda$ be the UV scale associated with the $(\text{particle size})^{-1}$ and the bare coupling $\lambda_\text{b}\equiv \frac{\Lambda^{2-d}}{m}\lambda$ . The dimensionless coupling $\lambda$ is subject to a 1-loop exact renormalization, which admits two fixed points $\lambda=0$ and $\lambda=\lambda_\text{f}<0$ when $4>d>2$. The bulk Schrödinger symmetry $Sch(d)$ is preserved at the fixed points, enabling us to study the impurity in a DCFT framework. In the next part, we first address the $\lambda=0$ fixed point, where the bulk theory is Gaussian. We then turn to discuss the $\lambda=\lambda_\text{f}$ fixed point of the unitary Fermi gas subsequently.

\subsection{Perturbative fixed points at $d=2+\Bar{\epsilon}$}

We first show that the s-wave resonance fixed point in \eqref{eq_bulk-to-defect EoM} and \eqref{eq_exact defect propagator} can be obtained by fine-tuning a contact interaction $\mu_\text{b}=\delta^d(x)\frac{\Lambda^{2-d}}{m}\hat{\mu}$. Physically, $\hat{\mu}$ can be understood as a local spike of the chemical potential, which can be engineered by external fields such as laser beams. In cases where the impurity is induced by heavy atoms, $\hat{\mu}$ marks the two-body interaction between bulk particles and the impurity. We remark that, experimentally, one candidate for such interactions is the polarization potential \cite{PhysRevLett.102.230402,zipkes2010trapped}, where the long-distance potential $V(r\to \infty)$ decays faster than $O(r^{-2})$.

\begin{figure}[htb]
\centering
  \includegraphics[width=.8\textwidth ]{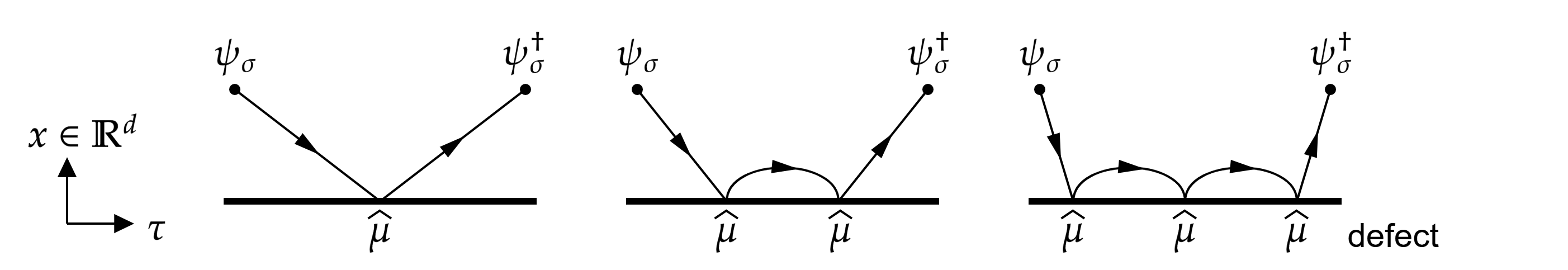 }
  \caption{\label{pic_defect chemical potential} Defect Feynman diagrams induced by the contact interaction $\hat{\mu}$. 
Throughout this paper, we use circles for inserted operators, bold lines for the defect (impurity) location, and directional lines for the free field propagators \eqref{eq_free field propagator}. }
\end{figure}

Dimensional analysis suggests that the renormalization group flow of $\hat{\mu}$ is perturbative around $d=2$, which 
motivates us to take $d=2+\Bar{\epsilon}$. We note that the defect bilinear deformation $\int_{x=0}d\tau \psi^\dagger_\sigma \psi_\sigma$ is of the double-trace type. It therefore leads to a one-loop exact renormalization group flow. Indeed, the defect ladder diagrams by powers of $\hat{\mu}$ are as in figure \ref{pic_defect chemical potential}, from which we obtain the beta function:
\begin{equation}
\label{eq_d=2 defect coupling RG}
-\beta(\hat{\mu})=-\bar{\epsilon} \hat{\mu}-\frac{\hat{\mu}^2}{\pi }+O\left(\bar{\epsilon}^3\right)~,
\end{equation}
and the nontrivial fixed point value reads $\hat{\mu}_\text{f}=-\pi \bar{\epsilon}+O\left(\bar{\epsilon}^2\right)$. For $d>2$, the fixed point is multicritical, and its negativity $\hat{\mu}_\text{f}<0$ signifies the interaction between the impurity and particles to be attractive. For $d<2$, the fixed point is stable in infrared with the impurity-particle interaction being repulsive. We remark that attractive defects in relativistic free theories usually lead to an ``unhealthy" DCFT \footnote{Such ``unhealthiness" can include instability, run-away behavior, and violation of unitarity.}, due to anti-particles and second quantization. However, such defects can exist in non-relativistic Schrödinger theories, where particle numbers are conserved. 

The theory remains Gaussian with the bilinear defect deformation. Therefore, the complete defect spectrum at $\hat{\mu}=\hat{\mu}_\text{f}$ follows the Wick theorem. In addition to one-body operators ($\Delta(\hat{\psi}^\dagger_{\sigma,l\geq 1})=\frac{d}{2}+l$ and $\Delta(\hat{\psi}^\dagger_{\sigma,0})=2-\frac{d}{2}$) addressed in section \ref{sec_DCFT Generalities}, we also note the lowest-lying two body operators with different spins: 
\begin{equation}
\label{eq_free theory defect spectrum}
\begin{aligned}
\Delta(\hat{\Phi}^\dagger_{0})=&2\Delta(\hat{\psi}^\dagger_{\sigma,0})=4-d~,\\
\Delta(\hat{\Phi}^\dagger_{l\geq 1})=&\Delta(\hat{\psi}^\dagger_{\sigma,l})+\Delta(\hat{\psi}^\dagger_{\sigma,0})=2+l~.\\
\end{aligned}
\end{equation}
The generalization of \eqref{eq_free theory defect spectrum} to many-body operators is straightforward. Under the state/operator correspondence, the Wick theorem states that many-body wave functions are the product (Slater determinant) of one-body functions. From the scaling dimensions \eqref{eq_free theory defect spectrum}, we learn that the ground state of a spinning two-body state is where one particle forms a bound state while the other spins around the impurity.

\subsection{Impurity interactions}

Even though the impurity spectrum at the fixed point $\hat{\mu}=\hat{\mu}_\text{f}$ of 
equation \eqref{eq_d=2 defect coupling RG} is exactly solvable, in realistic setups various non-linear interaction terms can arise on the defect. We would like to understand if the DCFT \eqref{eq_free theory defect spectrum} is infrared stable against such interactions. In \eqref{eq_free theory defect spectrum}, 
note that the lowest-lying two body operator has $\Delta(\hat{\Phi}_0^\dagger)=4-d$. It follows from the non-relativistic normal ordering that $\Delta(\hat{\Phi}_0^\dagger\hat{\Phi}_0)=2 \Delta(\hat{\Phi}_0^\dagger)$, and the defect deformation $ \int_{x=0}d\tau \hat{\Phi}_0^\dagger\hat{\Phi}_0$ is close to marginality when $d\approx 3$. We first fine-tune the defect two-body interaction $\hat{\mu}$ to attain the fixed point value $\hat\mu_\text{f}$ and denote the corresponding resonance Lagrangian density $\mathcal{L}_{\text{res}}$. We then perturb it as:
\begin{equation}\label{eq_impurity_interaction_term_action}
\mathcal{L}_{\text{res}} \to \mathcal{L}_{\text{res}}+\delta^d(x)\frac{\Lambda^{2d-6}}{m}\hat{\lambda}\hat{\Phi}_0^\dagger\hat{\Phi}_0~.
\end{equation}
From a perturbative analysis in $d=3+\Tilde{\epsilon}$, the beta function of the defect quartic coupling $\hat{\lambda}$ reads:
\begin{equation}
\label{eq_defect interaction flow}
-\beta(\hat{\lambda})=2\Tilde{\epsilon}\hat{\lambda}-\frac{\hat{\lambda}^2}{4\pi^3}+O\left(\Tilde{\epsilon}^3\right)~.
\end{equation}
The RG flow is again one-loop exact. The above beta-function admits two fixed points, at $\hat{\lambda}=0$ and $\hat{\lambda}=\hat{\lambda}_\text{f}=8\pi^3\tilde{\epsilon}$. When $d<3$, the $\hat{\lambda}=0$ fixed point is infrared stable. The interacting fixed point $\hat{\lambda}=\hat{\lambda}_\text{f}$ flips its sign across $d=3$, and becomes infrared stable when $d>3$. The scenario is similar to $\hat{\mu}=\hat{\mu}_\text{f}$ fixed point across $d=2$, the negative quartic coupling $\hat{\lambda}=\hat{\lambda}_\text{f}$ marks resonances and is well-defined in $U(1)_M$-perserving systems. Note the renormalization flow \eqref{eq_defect interaction flow} at $d=3$ is logarithmic, similar to the relativistic $\text{QED}_{4}$.

By fine-tuning the defect two-body interaction, the fixed point $\hat{\mu}_\text{f}$ can be obtained when the one-body wave function is in resonance with the impurity. In the free Fermi gas, we found the impurity physics is different below and above critical dimension $d=3$: When $d<3$, non-linear interactions are suppressed at long distance, the many-body wave function remains linear, and critical exponents are as in \eqref{eq_free theory defect spectrum}. On the other hand when $d>3$, such interactions are important. At the fixed point $\hat{\lambda}=\hat{\lambda}_\text{f}>0$, the two body operator dimension reads $\Delta(\hat{\Phi}^\dagger_{0})=d-2>4-d$. From \eqref{eq_defect wave function}, it indicates the quartic interaction tames the impurity attraction and the many-body wave function is non-linear.

\section{Unitary Fermi Gas}
\label{sec_Unitary Fermi Gas}

We proceed to study the $\lambda=\lambda_\text{f}<0$ bulk fixed point of the model \eqref{eq_general fermi lagragian}. Note that $\lambda<0$ marks an attractive two-body interaction, and it describes the unitary fermi gas at the Feshbach resonance \cite{regal2004observation,zwierlein2004condensation}. In the bulk, the one-body operator remains unperturbed $\Delta(\psi_\sigma^\dagger)=\frac{d}{2}$, while the two-body operator yields $\Delta(\Phi^\dagger)=2$, following the one-loop exactness of the renormalization flow \cite{nikolic2007renormalization,nishida2007nonrelativistic}. When the impurity (defect) is absent, the trivial DCFT coincides with the bulk conformal data, such that 
\begin{equation}
\label{eq_UFG bulk data}
\Delta(\hat{\psi}_{\sigma,{l }}^\dagger)=\frac{d}{2}+l\text{, and }\Delta(\hat{\Phi}^\dagger_{l})=2+l~.
\end{equation}
As far as the one- and two-body operators considered, \eqref{eq_UFG bulk data} differs from \eqref{eq_free theory defect spectrum} in the $l=0$ component. We remark that since the theory is non-Gaussian at $\lambda=\lambda_\text{f}$, the many-body operator spectrum is in general different from the double-twist type \cite{lauria2021line} of equation \eqref{eq_free theory defect spectrum}. 

We engineer the defect by fine-tuning the impurity contact term $\mu_\text{b}=\delta^d(x)\frac{\Lambda^{2-d}}{m}\hat{\mu}$ as in section \ref{sec_free fermi gas}. 
The renormalization of $\hat{\mu}$ is determined from the one-body sector of the theory, 
as given in equation \eqref{eq_d=2 defect coupling RG} and figure \ref{pic_defect chemical potential}, even with the presence of the bulk quartic interaction $\lambda<0$. 
This agrees with the physical intuition that a single particle is oblivious to the many-body interactions. More concretely, we note that the normal-ordered quartic operator annihilates all single-particle states. It follows from the Lehmann–Symanzik–Zimmermann reduction that such vertices do not contribute to Feynman diagrams as in figure \ref{pic_defect chemical potential}. As discussed above, the one-body defect operators at $\hat{\mu}=\hat{\mu}_\text{f}$ are of the dimension $\Delta(\hat{\psi}_{\sigma,{l=0}}^\dagger)=2-\frac{d}{2}$ and $\Delta(\hat{\psi}_{\sigma,{l\geq 1}}^\dagger)=\frac{d}{2}+l$.

We would like to understand whether the impurity interactions are relevant at the $\hat{\mu}=\hat{\mu}_\text{f}$ resonance. 
We compute the scaling dimensions of two-body operators,  
among which $\Phi^\dagger_{0}$ is the lowest-lying one. Particularly interesting is the case of $d=3$ spatial dimensions,  the bulk fixed point \eqref{eq_general fermi lagragian} becomes strongly coupled. The strongly coupled regime can be approached by perturbative analysis around $d=2$ and around $d=4$ \cite{nishida2007fermi,nishida2007nonrelativistic,nishida2011unitary}. Below, we present the analysis for each of these two approaches.

\subsection{$d=2+\Bar{\epsilon}$}

\begin{figure}[htb]
\centering
\includegraphics[width=.8\textwidth ]{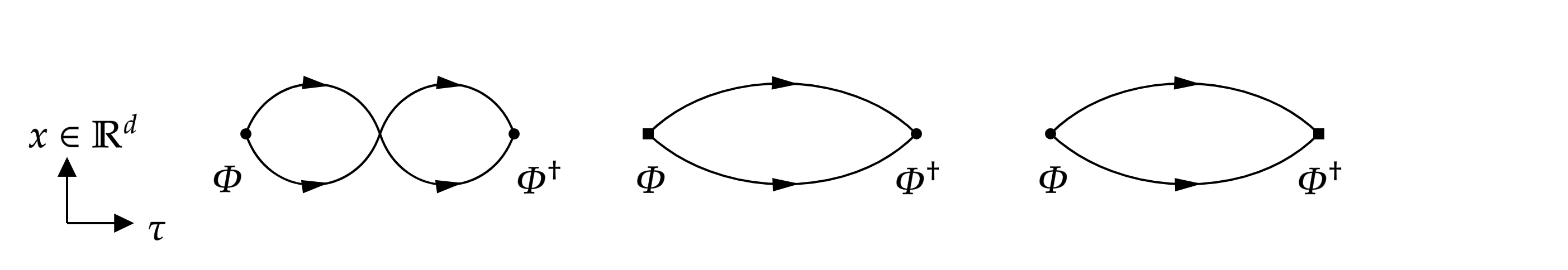 }
  \caption{\label{pic_d=2 bulk feynman diagram}$d=2+\Bar{\epsilon}$ bulk Feynman diagrams at $O(\Bar{\epsilon})$, where boxes marks counter terms. The corresponding integral is evaluated in appendix \ref{appsec_d=2 bulk 1-loop}.}
\end{figure}

The bulk coupling $\lambda_\text{b}$ in \eqref{eq_general fermi lagragian}  approaches marginality when $d\to 2$, and correspondingly, the correlation function of the composite operator $\Phi^\dagger$ is dominated by the tree-level propagators. In the Gross-Neveu representation, the bulk theory reads
\begin{equation}
\label{eq_d=2 bulk Largragian}
\mathcal{L}=\sum_{\sigma=\uparrow,\downarrow}\psi^\dagger_{\sigma}\left(\partial_\tau-\frac{\nabla^2}{2m}\right)\psi_\sigma-\frac{1}{\lambda_\text{b}}\Phi^\dagger\Phi+\left(\Phi^\dagger \psi_\downarrow \psi_\uparrow+\text{h.c.}\right)~,
\end{equation}
where $\Phi$'s equation of motion yields $\Phi^\dagger=\lambda_\text{b}\psi_\uparrow^\dagger\psi_\downarrow^\dagger$. With $\lambda_\text{b}\equiv \frac{\Lambda^{2-d}}{m}\lambda$ and the assumption $\lambda=O\left(\bar{\epsilon}\right)$, 
the renormalization counter term $\delta \mathcal{L}=\frac{1}{\lambda_\text{b}}\Phi^\dagger \Phi$ is required for a valid perturbative expansion. The one-loop correction to the two-point function $\langle\Phi \Phi^\dagger \rangle$ is in figure \ref{pic_d=2 bulk feynman diagram}, from which we obtain the renormalization function: 
\begin{equation}\label{eq_d=2_bulk_RG_flow}
   - \beta(\lambda)= -\bar{\epsilon}\lambda-\frac{\lambda^2}{2\pi}+O(\bar\epsilon^3).
\end{equation}
As we noted previously, the RG flow is one-loop exact. The beta-function \eqref{eq_d=2_bulk_RG_flow} admits the fixed point $\lambda=\lambda_\text{f}=-2\pi\bar{\epsilon}+O\left(\bar{\epsilon}\right)$ that marks the unitary Fermi gas \cite{nikolic2007renormalization}.

\begin{figure}[htb]
\centering
  \includegraphics[width=.8\textwidth ]{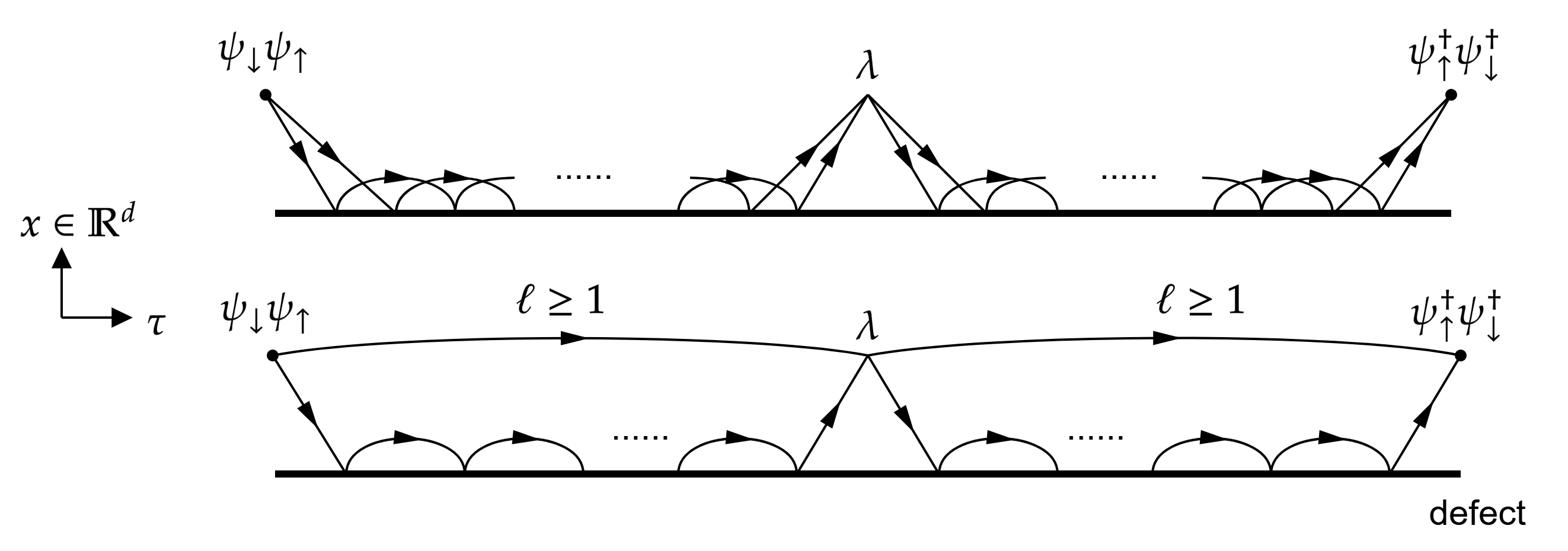}
  \caption{\label{pic_d=2 defect 1-loop}$d=2+\Bar{\epsilon}$
  Feynman diagrams contributing to the $\hat{\Phi}^\dagger_{l\in \mathbb{N}}$ anomalous dimension at $O(\Bar{\epsilon})$. The corresponding integral is evaluated in appendix \ref{appsec_d=2 defect anomalous dimension}.}
\end{figure}

As we remarked previously, the scaling dimension of defect one-body operators does not receive perturbative correction from the bulk coupling $\lambda_\text{b}$. On the other hand, the defect two-body operators in \eqref{eq_free theory defect spectrum} do receive corrections as in the Feynman diagrams \ref{pic_d=2 defect 1-loop}. At the fixed point $\lambda=\lambda_\text{f}$, we find:
\begin{equation}
\label{eq_d=2 int defect dimension}
\begin{aligned}
\Delta(\hat{\Phi}^\dagger_{0})=&4-d+\frac{\lambda_\text{f}}{2\pi}+O\left(\lambda_\text{f}^2\right)=2-2\bar{\epsilon}+O\left(\bar{\epsilon}^2\right)~,\\
\Delta(\hat{\Phi}^\dagger_{l \geq 1})=&2+l+\frac{\lambda_\text{f}}{2^l \pi}+O\left(\lambda_\text{f}^2\right)=2+l-\frac{\bar{\epsilon}}{2^{l-1}}+O\left(\bar{\epsilon}^2\right)~.
\end{aligned}
\end{equation}
In the state/operator correspondence picture, the scaling dimensions \eqref{eq_d=2 int defect dimension} indicate that the energy level is perturbed by the attractive interaction between the bounded particle and the spinning one. Note that the quantum correction is exponentially suppressed at large spin $l$, and a Regge limit emerges \cite{alday2007comments,komargodski2013convexity,cuomo2023giant}. Finally, equation \eqref{eq_d=2 int defect dimension} suggests that the critical spatial dimension where $ \int_{x=0}d\tau \hat{\Phi}_0^\dagger\hat{\Phi}_0$ becomes relevant is $d\approx 2.5$.

\subsection{$d=4-\epsilon$}

\begin{figure}[htb]
\centering
\includegraphics[width=.8\textwidth ]{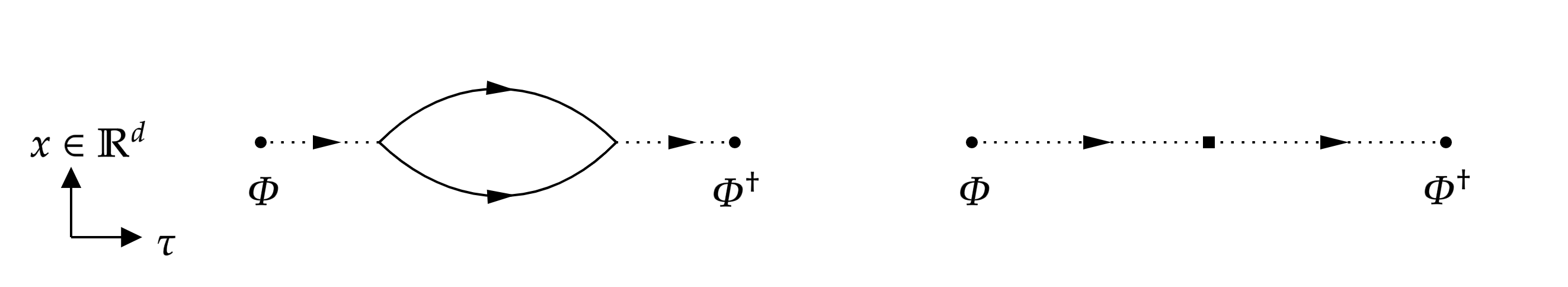}
  \caption{\label{pic_d=4 bulk feynman diagram}$d=4-\epsilon$ bulk Feynman diagrams at $O(\epsilon)$. The corresponding integral is evaluated in appendix \ref{appsec_d=4 bulk 1-loop}.}
\end{figure}

When the spatial dimension approaches $d=4$ from below, the propagator of the two-body operator $\Phi^\dagger$ is dominated by infinite ladder diagrams. We note that $\Delta(\Phi^\dagger)=2$ is close to the unitarity bound \cite{pal2018unitarity} of $\frac{d}{2}$, suggesting such diagrams admit a perturbed free field propagator representation. Indeed \cite{nishida2007fermi, Nussinov:2006zz}, the Gross-Neveu representation of the unitary Fermi gas \eqref{eq_general fermi lagragian} in $d=4-\epsilon$ reads
\begin{equation}
\label{eq_d=4 bare Largragian}
\mathcal{L}=\sum_{\sigma=\uparrow,\downarrow}\psi^\dagger_{\sigma}\left(\partial_\tau-\frac{\nabla^2}{2m}\right)\psi_\sigma+\Phi^\dagger\left(\partial_\tau-\frac{\nabla^2}{4m} \right)\Phi+\left(g_\text{b}\Phi^\dagger \psi_\downarrow \psi_\uparrow+\text{h.c.}\right)~.
\end{equation}
Notably, a counter term $\delta \mathcal{L}=-\Phi^\dagger(\partial_\tau-\frac{\nabla^2}{4m} )\Phi $ is necessary to avoid repeated counting of the infinite ladder diagrams. The Yukawa coupling $g_\text{b}\equiv \Lambda^{\epsilon/2}g/m$ marks a relevant deformation, and its renormalization can be obtained from the diagrams in figure \ref{pic_d=4 bulk feynman diagram}.
We find the beta function
\begin{equation}
\label{eq_d=4 bulk coupling RG}
-\beta(|g|^2)=\epsilon |g|^2-\frac{|g|^4}{8\pi^2}+O\left(\epsilon^3\right)~,
\end{equation}
with the fixed point value $|g|^2_\text{f}=8\pi^2 \epsilon+O\left(\epsilon^2\right)$ \cite{nishida2007nonrelativistic}.

\begin{figure}[thb]
\centering
\includegraphics[width=.8\textwidth ]{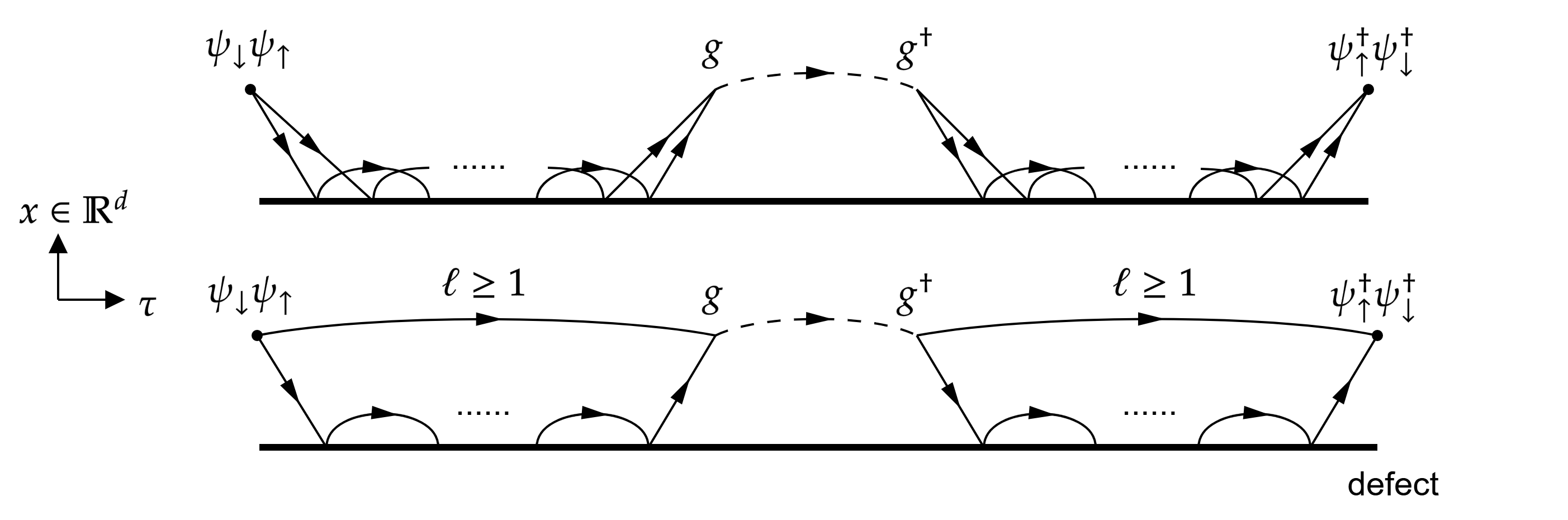}
  \caption{\label{pic_d=4 defect 1-loop}$d=4-\epsilon$ defect Feynman diagrams of the $\hat{\Phi}^\dagger_{l\in \mathbb{N}}$ anomalous dimension at $O(\epsilon)$. The corresponding integral is evaluated in appendix \ref{appsec_d=4 defect anomalous dimension}.}
\end{figure}

The defect operator corrections follow a similar Feynman diagram calculation as the $d=2+\bar{\epsilon}$ case, which is illustrated in figure \ref{pic_d=4 defect 1-loop}. At the fixed point $|g|=|g|_\text{f}$, we find:
\begin{equation}
\label{eq_d=4 defect dimension spin 0}
\Delta(\hat{\Phi}^\dagger_{0})=4-d+\frac{2|g|^2_\text{f}}{3\pi^2}+O\left(|g|_\text{f}^4\right)=\frac{19}{3}\epsilon+O\left(\epsilon^2\right)~.
\end{equation}
Note that $\Delta(\hat{\Phi}^\dagger_{0})$ marks the energy level of two fermions bounded to the impurity, and it is raised by quantum corrections. On the other hand, we find for spinning operators:
\begin{equation}
\label{eq_d=4 defect dimension spin l}
\Delta(\hat{\Phi}^\dagger_{l\geq 1})=2+l+O\left(\epsilon|g|_\text{f}^2, |g|_\text{f}^4\right)=2+l+O\left(\epsilon^2\right)~,
\end{equation}
such that they remain unperturbed at $O(\epsilon)$. We remark that \eqref{eq_d=4 defect dimension spin l} implies the decoupling between the spinning particle and the bounded one at the limit $d\to 4$. Equations \eqref{eq_d=4 defect dimension spin 0} and \eqref{eq_d=4 defect dimension spin l} 
suggest a physical picture different from that in \eqref{eq_d=2 int defect dimension}, and we accordingly identify the critical dimension $d\approx 3.84$.

\section{Discussion and Outlook}

\begin{figure}[htb]
\centering
\includegraphics[width=.5\textwidth ]{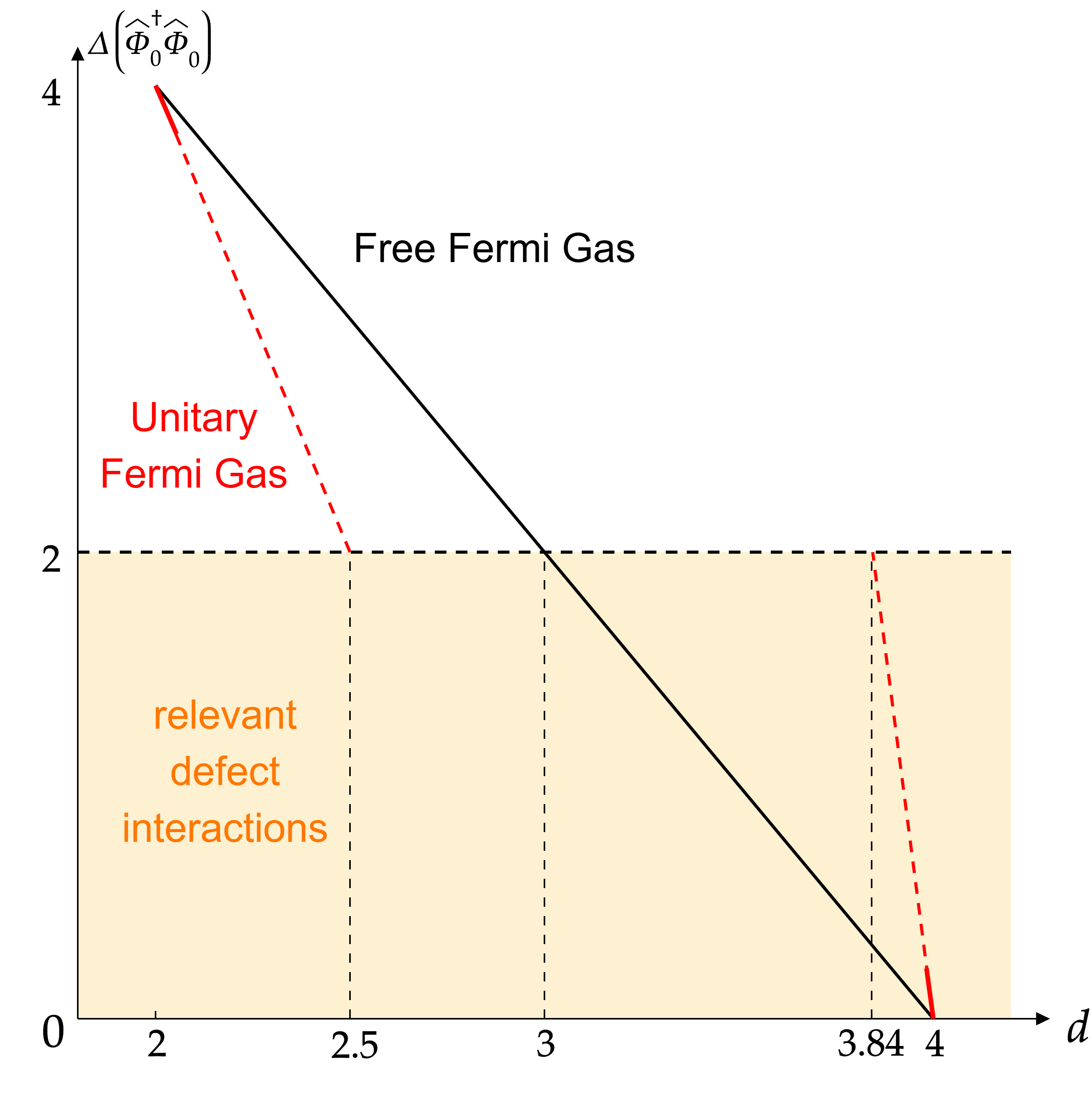}
  \caption{\label{pic_phase diagram}Scaling dimensions of the defect interaction term $\hat{\Phi}^\dagger_0\hat{\Phi}_0$. The black/red solid line marks the scaling dimensions obtained in the free/unitary Fermi gas. The red dashed lines are extensions of the linear dependence of $\Delta(\hat{\Phi}_0^\dagger\hat{\Phi}_0)$ in \eqref{eq_d=2 int defect dimension} and \eqref{eq_d=4 defect dimension spin 0}. }
\end{figure}

With the presence of an impurity, we have analytically solved the one-body sector for a generic local Schrödinger field theory. We have obtained the exact impurity spectrum at s-wave resonance for the free Fermi gas. For the interacting unitary Fermi gas, corresponding predictions are made around spatial dimensions $d=2$ and $d=4$. A summary of our results for the free and unitary Fermi gas is given by the phase diagram \ref{pic_phase diagram}. 

Perturbative analysis has suggested no definitive answer to whether the interaction $\hat{\Phi}_0^\dagger\hat{\Phi}_0$ is relevant on a resonant impurity of the $d=3$ unitary Fermi gas. This unresolved stability problem is related to the Efimov effect \cite{efimov1970weakly,efimov1973energy,ovchinnikov1979number, braaten2006universality}, where the impurity is viewed as a heavy bosonic particle. In such a picture, two light fermions scatter off the impurity through the three-body interaction $\hat{\Phi}_0^\dagger\hat{\Phi}_0$. It would be alluring to understand if a comparison between different methods toward the same defect universality class can be drawn. We comment that as in \cite{nishida2007nonrelativistic}, our analysis can be straightforwardly extended to the defect spectrum of higher $n$-particle states.

Exploring experimental configurations to test our theory is clearly desirable. Even though the ambient bulk criticality can be approached by fine-tuning the external magnetic field \cite{regal2004observation,zwierlein2004condensation} in atomic gas systems, we are not aware of feasible methods that can attain the impurity s-wave resonance. A plausible candidate is the near-conformal setup, where a heavy particle of large scattering length is involved in the scattering process. Another platform suggested by the state/operator correspondence \eqref{eq_defect wave function} is the many-body states in harmonic traps \cite{werner2006unitary,werner2006unitaryap,busch1998two,gharashi2015one}. We expect appropriate trap deformations to be identified in the defect universality class that we studied and can be approached either numerically or experimentally.
 
We conclude by outlining future directions and open questions that are of theoretical interest concerning impurities in Schrödinger field theories:

\begin{itemize}

\item{$g$-theorem}: Conformal line defects in Schrödinger field theories preserve the same $SL(2,\mathbb{R})$ as they do in relativistic CFTs. It is of fundamental importance to find out whether a modified version or analog of the g-theorem \cite{cuomo2022renormalization} exists in anisotropic Lifshitz theories.

\item{Large charge sector}: The interacting spectrum of bulk operators with large charge quantum numbers can be extracted from a conformal superfluid EFT \cite{badel2019epsilon,cuomo2023giant, hellerman2024unitary}. It will be interesting to understand the EFT counterpart of the defect and extract the large charge spectrum.

\item{Anyonic models}: Wilson lines are important non-local observables in $(2+1)d$ Chern-Simons-matter theories. 
In anyonic models of the non-relativistic regime \cite{doroud2016superconformal,doroud2018conformal}, it would be desirable to understand the operator spectrum of $SL(2,\mathbb{R})$ Wilson lines.

\item{Spin impurity}: As it is closely related to the Kondo problem\cite{de1934electrical,kondo1964resistance}, spin impurities are line defects of great experimental interests (see e.g. \cite{cuomo2022spin, sachdev1999quantum, vojta2000quantum}).
An interesting open problem is to study the spin coupling in Fermi gases and their long-distance contributions.

\end{itemize}

\section*{Acknowledgments}
We thank Luca Delacretaz, Zohar Komargodski, Amit Sever, Dam Thanh Son, and Yunqin Zheng for various useful discussions. 
We are especially grateful to Gabriel Cuomo and Zohar Komargodski for providing comments on a preliminary version of this manuscript.
ARM is supported by the Simons Center for Geometry and Physics.
SZ is supported in part by the Simons Foundation grant 488657 (Simons Collaboration on the Non-Perturbative Bootstrap), the BSF grant no. 2018204 and NSF award number 2310283.

\appendix

\newpage

\section{Special Functions}

The  $d$-dimensional spherical harmonics $Y_{l,q}$ form a complete and orthogonal basis of square-integrable functions on $\Omega \in \mathbb{S}^{d-1}$, and they satisfy:
\begin{equation}
\label{appendix_spherical harmonics}
\begin{aligned}
&-\partial^2_{\mathbb{S}^{d-1}}Y_{l,q}(\Omega)=l(l+d-2)Y_{l,q}(\Omega)~,\\
&\int_{\mathbb{S}^{d-1}}d \Omega Y^*_{l,q}(\Omega)Y_{{l}',{q}'}(\Omega)=\delta_{l,{l}'}\delta_{q,{q}'}~.\\
\end{aligned}
\end{equation}
The volume of the $\mathbb{S}^{d-1}$ sphere and the degeneracy of the $l$-th spherical harmonic read:
\begin{equation}
\begin{aligned}
|\mathbb{S}^{d-1}|= & \frac{2 \pi ^{d/2}}{\Gamma \left(d/2\right)}~,\\
\text{deg}(d,l)= & (2l+d-2)\frac{\Gamma (l+d -2)}{\Gamma (l+1)\Gamma (d-1)}~.
\end{aligned}
\end{equation}
The $0$-th spherical harmonic $Y_{0}=1/\sqrt{|\mathbb{S}^{d-1}|}$ corresponds to the s-wave scattering, and is non-degenerate. 
%and we have 
Let $\cos{\theta}=\Omega\cdot {\Omega}'$ be the angle between two points on $\mathbb{S}^{d-1}$, the addition theorem states that 
\begin{equation}
C_l(\cos {\theta})=\frac{|\mathbb{S}^{d-1}|}{\text{deg}(d,l)}\sum_{q=1}^{\text{deg}(d,l)} Y^*_{l,q}(\Omega) Y_{l,q}({\Omega}')~,
\end{equation}
where $C_l$ is the $l$-th Gegenbauer polynomial of degree $\frac{d-2}{2}$, and we have chose the normalization such that $C_{l\in\mathbb{N}}(1)=1$. Specifically, it follows from the Rodrigues formula that 
\begin{equation}
C_l(t)=\frac{(1-t^2)^{\frac{3-d}{2}}}{(l+\frac{d-3}{2})_l}\left(-\frac{d}{2dt}\right)^l(1-t^2)^{l+\frac{d-3}{2}}~,
\end{equation}
where $C_l(1)=1$. The Gegenbauer polynomials form a complete and orthogonal basis of piecewise continuous functions on $[-1,1]$, such that:
\begin{equation}
\int_{-1}^{1}C_l(t)C_{{l}'}(t)(1-t^2)^{\frac{d-3}{2}}dt=\frac{\delta_{l,{l}'}|\mathbb{S}^{d-1}|}{\text{deg}(d,l)|\mathbb{S}^{d-2}|}~.
\end{equation}

Two integrals of Bessel functions were also of use throughout this work. We note the special case of the Weber-Schafheitlin integral \cite{kellendonk2009weber}: 
\begin{equation}
\label{appendix_Weber-Schafheitlin integral}
\int_{0}^{+\infty} kJ_{\nu}(kr_1)J_{\nu}(kr_2)dk=\frac{1}{r_1} \delta(r_1-r_2)~.
\end{equation}
\eqref{appendix_Weber-Schafheitlin integral} as a distribution on $\mathbb{R}^+$ is convergent for index $\nu >-1$, in agreement with the defect unitarity bound. Another important integral for defect correlation functions is given by:
\begin{equation}
\int_{0}^{+\infty} ke^{-pk^2}J_{\nu}(kr_1)J_{\nu}(kr_2)dk=\frac{1}{2p}e^{-\frac{\left(r_1^2+r_2^2\right)}{4p }} I_\nu\left(\frac{ r_1 r_2}{2p}\right)~,
\end{equation}
where $p>0$.

\section{Bulk Feynman Diagrams}

In this and the following appendix, we study the integrals of Feynman diagrams discussed in the main text. For simplicity, we apply the shorthand notations $\tau_{ij}\equiv\tau_i-\tau_j$, $x_{ij}\equiv x_i-x_j$, as in the main text.

\subsection{$d=2+\bar{\epsilon}$}
\label{appsec_d=2 bulk 1-loop}

We evaluate the one-loop integral of the bulk quartic vertex as in equation \eqref{eq_d=2 bulk Largragian}: 
\begin{equation}
\label{appendix_d=2 bulk 1-loop}
\begin{aligned}
&\int d\tau_2 d^dx_2 G^2_{m} (\tau_{12},x_{12}) G^2_{m}(\tau_{23},x_{23})\\
=&\frac{\Theta(\tau_{13})}{(2\tau_{13})^{\frac{d}{2}}}\left(\frac{m}{2\pi}\right)^{\frac{3d}{2}} e^{-\frac{m x_{13}^2}{\tau_{13}}}\int_{\tau_1>\tau_2>\tau_3}\frac{d\tau_2}{(\tau_{12}\tau_{23})^{\frac{d}{2}}}\\
=&\frac{\sqrt{\pi} 2^{\frac{d}{2}-1}\Gamma \left(1-\frac{d}{2}\right)}{\Gamma
   \left(\frac{3}{2}-\frac{d}{2}\right)} \left(\frac{m}{2\pi}\right)^{\frac{3d}{2}} \frac{\Theta(\tau_{13})}{\tau _{13}^{\frac{3 d}{2}-1}}e^{-\frac{m x_{13}^2}{\tau_{13}}}\\
=& \left(\frac{m}{2\pi}\right)^{3}\frac{\Theta(\tau_{13})}{\tau _{13}^{2+\bar{\epsilon}}}e^{-\frac{m x_{13}^2}{\tau_{13}}}\left[-\frac{2}{\bar{\epsilon}}+\ln{\tau_{13}}-3\ln \frac{m}{2\pi}+O\left(\bar{\epsilon}\right)\right]~.
\end{aligned}
\end{equation}
The tree-level propagator of the composite $\Phi$ field is simply the delta function. From \eqref{appendix_d=2 bulk 1-loop}, one can obtain the RG function of the bulk coupling $\lambda$ as in \eqref{eq_d=2_bulk_RG_flow} and correspondingly anomalous dimensions of bulk operators to the precision of $O(\bar{\epsilon})$. 

\subsection{$d=4-\epsilon$}
\label{appsec_d=4 bulk 1-loop}

We elaborate on the loop integrals of the bulk cubic interaction in \eqref{eq_d=4 bare Largragian}. At the one-loop level, the Feynman diagram is 
given in figure \ref{pic_d=4 bulk feynman diagram} and the corresponding integral reads:
\begin{equation}
\label{appendix_d=4 bulk 1-loop}
\begin{aligned}
&\int \prod_{i=2}^3 d\tau_i d^dx_i G_{2m} (\tau_{12},x_{12}) G^2_{m}(\tau_{23},x_{23})G_{2m}(\tau_{34},x_{34})\\
=&\left(\frac{m}{2\pi}\right)^d\frac{\Theta(\tau_{14})}{\tau_{14}^{d/2}}e^{-\frac{mx_{14}^2}{\tau_{14}}}\int_{\tau_1>\tau_2>\tau_3>\tau_4}\frac{d\tau_2 d \tau_3}{(\tau_2-\tau_3)^{d/2}}\\
=&\frac{(m/\pi)^d}{(d-2)(d-4)}\frac{\Theta(\tau_{14})}{(2\tau_{14})^{d-2}}e^{-\frac{mx_{14}^2}{\tau_{14}}}\\
=&\left(\frac{m}{\pi}\right)^4\frac{\Theta(\tau_{14})}{(\tau_{14})^2}e^{-\frac{mx_{14}^2}{\tau_{14}}}\left[-\frac{1}{8\epsilon}-\frac{1}{8} \ln \frac{2\pi  \tau_{14}}{m} -1+O\left(\epsilon\right)\right]~.
\end{aligned}
\end{equation}
At the two-loop level, the corresponding integral 
%of concern 
reads
\begin{equation}
\label{appendix_d=4 bulk 2-loop}
\begin{aligned}
&\int \prod_{i=2}^5 d\tau_i d^dx_i G_{2m} (\tau_{12},x_{12}) G^2_{m}(\tau_{23},x_{23})G_{2m}(\tau_{34},x_{34})G^2_{m}(\tau_{45},x_{45})G_{2m}(\tau_{56},x_{56})\\
=&\left(\frac{m}{2\pi}\right)^{3d/2}\frac{\Theta(\tau_{16})}{(2\tau_{16})^{d/2}} e^{-\frac{m x_{16}^2}{\tau_{16} }}\int_{\tau_1>\tau_2>\tau_3>\tau_4>\tau_5>\tau_6}\frac{d\tau_2 d \tau_3d\tau_4d\tau_5}{(\tau_2-\tau_3)^{d/2}(\tau_4-\tau_5)^{d/2}}\\
=&\left(\frac{m}{2\pi}\right)^{3d/2}\frac{\Theta(\tau_{16})}{(\tau_{16})^{\frac{3d}{2}-4}} e^{-\frac{m x_{16}^2}{\tau_{16} }}\frac{\pi ^{3/2} 2^{\frac{d}{2}-2} }{(d-2)(d-4)\Gamma \left(\frac{5}{2}-\frac{d}{2}\right) \Gamma
   \left(\frac{d}{2}\right)\sin \left(\frac{\pi  d}{2}\right)}\\
=&\left(\frac{m}{\pi}\right)^{6}\frac{\Theta(\tau_{16})}{(\tau_{16})^{2}} e^{-\frac{m x_{16}^2}{\tau_{16} }}\left[\frac{1}{64\epsilon^2}+\frac{1}{64\epsilon}\left(1+\frac{3}{2}\ln \frac{2\pi \tau_{16}}{m}\right)+O\left(\epsilon^0\right)\right]~.
\end{aligned}
\end{equation}
From \eqref{appendix_d=4 bulk 1-loop} and \eqref{appendix_d=4 bulk 2-loop}, one can obtain the RG function of the bulk coupling $|g|^2$ as in \eqref{eq_d=4 bulk coupling RG} and the anomalous dimensions of bulk operators to the precision of $O(\epsilon)$. 

\section{Defect Feynman Diagrams}

When defect contributions are involved, it is useful to decompose the free field propagators according to:
\begin{equation}
\begin{aligned}
G_m(\tau,x_{12})=&\Theta(\tau)\frac{(m/\tau)e^{-\frac{m(r_1^2+r_2^2)}{2\tau}}}{(r_1 r_2)^{\frac{d}{2}-1}}\sum_{l\in \mathbb{N}}\sum_{q=1}^{\text{deg}(d,l)}Y_{l,q}(\Omega_1)Y^*_{l,q}(\Omega_2) I_{l+\frac{d}{2}-1}\left(\frac{mr_1r_2}{\tau}\right)\\
=&\Theta(\tau)\frac{(m/\tau)^{d/2}e^{-\frac{m(r_1^2+r_2^2)}{2\tau}}}{|\mathbb{S}^{d-1}|}\sum_{l\in \mathbb{N}}\frac{\text{deg}(d,l)}{\Gamma(\frac{d}{2}+l)} C_l(\cos\theta_{12})\left(\frac{mr_1r_2}{\tau}\right)^{l}\left[1+O\left(r_1^2,r_2^2\right)\right]~,\\
D_m(\tau,r_1,r_2)=&\Theta(\tau)\frac{(m/\tau) e^{-\frac{m \left(r_1^2+r_2^2\right)}{2 \tau }}
   }{ |\mathbb{S}^{d-1}|(r_1r_2)^{\frac{d}{2}-1}}\left[I_{1-\frac{d}{2}}\left(\frac{m r_1 r_2}{\tau
   }\right)-I_{\frac{d}{2}-1}\left(\frac{m r_1 r_2}{\tau }\right)\right]~,
\end{aligned}
\end{equation}
where $\cos \theta_{12}\equiv \Omega_1 \cdot \Omega_2$ marks the relative angle between the two external spatial points. For two-body propagators, the tree-level propagator follows the Wick theorem and it reads
\begin{equation}
\label{appendix_defect tree level}
\mathtoolsset{multlined-width=0.9\displaywidth}
\begin{multlined}
\phantom{=}\left[G_{m}(\tau,x_{12})+D_m(\tau,r_1,r_2)\right]^2\\
=\Theta(\tau)\frac{(m/\tau)^2 e^{-\frac{m \left(r_1^2+r_2^2\right)}{ \tau }} 
   }{ |\mathbb{S}^{d-1}|^2(r_1r_2)^{d-2}}\left\{\left[I_{1-\frac{d}{2}}\left(\frac{m r_1 r_2}{\tau
   }\right)-I_{\frac{d}{2}-1}\left(\frac{m r_1 r_2}{\tau }\right)\right]^2\right.\hfill \\
+2|\mathbb{S}^{d-1}|\sum_{l\in \mathbb{N}}\sum_{q=1}^{\text{deg}(d,l)}Y_{l,q}(\Omega_1)Y^*_{l,q}(\Omega_2) \left[\frac{|\mathbb{S}^{d-1}| }{(4\pi)^{\frac{d}{2}}}\left(\frac{mr_1r_2}{\tau}\right)^{\frac{d}{2}-1}I_{l+\frac{d}{2}-1}\left(\frac{2mr_1r_2}{\tau}\right)\right.\\
   \hfill \left.\left. +I_{l+\frac{d}{2}-1}\left(\frac{mr_1r_2}{\tau}\right)\left(I_{1-\frac{d}{2}}\left(\frac{m r_1 r_2}{\tau
   }\right)-I_{\frac{d}{2}-1}\left(\frac{m r_1 r_2}{\tau }\right)\right)\right]\right\}\\
   =\Theta(\tau)\frac{(m/\tau)^2 e^{-\frac{m \left(r_1^2+r_2^2\right)}{ \tau }} 
   }{ \left(|\mathbb{S}^{d-1}|\Gamma(2-\frac{d}{2})\right)^2(r_1r_2)^{d-2}}\left\{ \left[\left(\frac{mr_1r_2}{2\tau}\right)^{2-d}+O\left((r_1r_2)^0\right)\right]\right.\hfill\\
   \hfill \left.+2 \Gamma(2-\frac{d}{2}) \sum_{l\geq 1}\frac{\text{deg}(d,l)C_l(\cos{\theta_{12}})}{\Gamma(\frac{d}{2}+l)}\left[\left(\frac{mr_1r_2}{2\tau}\right)^l+O\left((r_1r_2)^{l+d-2}\right)\right]\right\}~.
\end{multlined}
\end{equation}
The lowest-lying defect two-body operators from \eqref{appendix_defect tree level} are of the dimensions $\Delta(\hat{\Phi}^\dagger_0)=4-d$ and $\Delta(\hat{\Phi}^\dagger_{l\geq1})=2+l$, in agreement with \eqref{eq_free theory defect spectrum}.

\subsection{$d=2+\bar{\epsilon}$}
\label{appsec_d=2 defect anomalous dimension}

The bulk quartic vertex contributes to the following one-loop integral:

\begin{equation}
\label{appendix_d=2 defect anomalous dimension part 1}
\mathtoolsset{multlined-width=0.9\displaywidth}
\begin{multlined}
    \phantom{=}\int  d\tau_2 d^dx_2 [G_{m}(\tau_{12},x_{12})+D_m(\tau_{12},r_1,r_2)]^2 [G_{m}(\tau_{23},x_{23})+D_m(\tau_{23},r_2,r_3)]^2\\
    =\frac{\Theta(\tau_{13})(m/\tau_{13})^2  
   }{ \left(|\mathbb{S}^{d-1}|\Gamma(2-\frac{d}{2})\right)^2(r_1r_3)^{d-2}}\left\{\frac{4^{d-2}  m^{4-d} \tau _{13}^2 }{|\mathbb{S}^{d-1}| (\Gamma
   \left(2-\frac{d}{2}\right))^2\left(m r_1
   r_3\right)^{d-2}}\right. \hfill\\
   \times \int_{\tau_1>\tau_2>\tau_3}\frac{d\tau_2}{(\tau_{12}\tau_{23})^{4-d}}\int dr_2 e^{-m\left(\frac{r_1^2+r_2^2}{\tau _{12}}+\frac{r_2^2+r_3^2}{\tau _{23}}\right)}\left[r_2^{7-3 d} +O\left(r_1^{d-2},r_3^{d-2}\right)\right]\\
   +\frac{(2m \tau_{13})^2}{|\mathbb{S}^{d-1}|}\sum_{l\geq 1}\frac{\deg(d,l)C_l(\cos{\theta_{13}})}{(\Gamma \left(\frac{d}{2}+l\right))^2}\left(\frac{m^2r_1r_3}{4}\right)^l\\
   \left.\times\int_{\tau_1>\tau_2>\tau_3}\frac{d\tau_2}{(\tau_{12}\tau_{23})^{2+l}}\int dr_2 e^{-m\left(\frac{r_1^2+r_2^2}{\tau _{12}}+\frac{r_2^2+r_3^2}{\tau _{23}}\right)}\left[r_2^{3-d+2l} +O\left(r_1^{d-2},r_3^{d-2}\right)\right]\right\}\\
   =\frac{\Theta(\tau_{13})(m/\tau_{13})^2  
   }{ \left(|\mathbb{S}^{d-1}|\Gamma(2-\frac{d}{2})\right)^2(r_1r_3)^{d-2}}\left\{ \frac{ m^{d/2} \Gamma
   \left(4-\frac{3 d}{2}\right) \tau _{13}^{\frac{3 d}{2}-2}}{2^{5-2d}|\mathbb{S}^{d-1}| ( \Gamma
   \left(2-\frac{d}{2}\right))^2\left(m r_1
   r_3\right)^{d-2}}\right.\hfill\\
\left.+\frac{2 (m \tau_{13})^{\frac{d}{2}}}{|\mathbb{S}^{d-1}|}\sum_{l\geq 1}\frac{\deg(d,l) C_l(\cos{\theta_{13}})\Gamma
   \left(l+2-\frac{d}{2}\right)}{(\Gamma \left(\frac{d}{2}+l\right))^2}\left(\frac{mr_1r_3}{4\tau_{13}}\right)^l\right\}\\
\times\left[\int_{\tau_1>\tau_2>\tau_3}\frac{d\tau_2 }{(\tau_{12}\tau_{23})^{\frac{d}{2}}}e^{-m\left(\frac{r_1^2}{\tau _{12}}+\frac{r_3^2}{\tau _{23}}\right)}+O\left(r_1^{d-2},r_3^{d-2}\right)\right]~.
\end{multlined}
\end{equation}
Note that the $\int d\tau_2$ integral in the last line of \eqref{appendix_d=2 defect anomalous dimension part 1} is UV finite:
\begin{equation}
\label{appendix_d=2 defect anomalous dimension part 2}
    \begin{aligned}
        \int_{\tau_1>\tau_2>\tau_3}\frac{d\tau_2 }{(\tau_{12}\tau_{23})^{\frac{d}{2}}}e^{-m\left(\frac{r_1^2}{\tau _{12}}+\frac{r_3^2}{\tau _{23}}\right)}=2\tau_{13}^{-\frac{d}{2}}\left(\ln{\frac{\tau_{13}}{mr_1r_3}}-\gamma_\text{E}+O\left(r_1^2,r_3^2,\epsilon\right)\right)~.
    \end{aligned}
\end{equation}
We remark that \eqref{appendix_d=2 defect anomalous dimension part 2} implies that no extra defect counterterms are needed, which is consistent with the fact that renormalization of the defect two-body interaction associated with equation \eqref{eq_d=2 defect coupling RG} is one-loop exact, irrespective of the bulk interactions. Plugging equation \eqref{appendix_d=2 defect anomalous dimension part 2} into \eqref{appendix_d=2 defect anomalous dimension part 1}, we obtain:
\begin{equation}
\label{appendix_d=2 defect anomalous dimension part 3}
\mathtoolsset{multlined-width=0.9\displaywidth}
\begin{multlined}
    \phantom{=}\int  d\tau_2 d^dx_2 [G_{m}(\tau_{12},x_{12})+D_m(\tau_{12},r_1,r_2)]^2 [G_{m}(\tau_{23},x_{23})+D_m(\tau_{23},r_2,r_3)]^2\\
    =\frac{\Theta(\tau_{13})(m/\tau_{13})^2  
   }{ \left(|\mathbb{S}^{d-1}|\Gamma(2-\frac{d}{2})\right)^2(r_1r_3)^{d-2}}\left\{ \left(\frac{mr_1r_3}{2\tau_{13}}\right)^{2-d}\right.\hfill\\
   \left. +4 \Gamma(2-\frac{d}{2}) \sum_{l\geq 1}\frac{\text{deg}(d,l)C_l(\cos{\theta_{12}})}{\Gamma(\frac{d}{2}+l)}\left(\frac{mr_1r_3}{4\tau_{13}}\right)^l\hfill\right\}\\
   \times \frac{m}{|\mathbb{S}^{d-1}|}\left(\ln{\frac{\tau_{13}}{mr_1r_3}}-\gamma_\text{E}+O\left(r_1^2,r_3^2,\epsilon\right)\right)~.
\end{multlined}
\end{equation}
From \eqref{appendix_defect tree level} and \eqref{appendix_d=2 defect anomalous dimension part 3}, one can find the corrected defect scaling dimensions as in \eqref{eq_d=2 int defect dimension}.

\subsection{$d=4-\epsilon$}
\label{appsec_d=4 defect anomalous dimension}

The bulk Yukawa coupling yields the following 1-loop integral
\begin{equation}
\label{appendix_d=4 defect anomalous dimension part 1}
\mathtoolsset{multlined-width=0.9\displaywidth}
\begin{multlined}
\int \prod_{i=2}^3 d\tau_i d^dx_i [G_{m}(\tau_{12},x_{12})+D_m(\tau_{12},r_1,r_2)]^2 G_{2m}(\tau_{23},x_{23})[G_{m}(\tau_{34},x_{34})+D_m(\tau_{34},r_3,r_4)]^2\\
    =\frac{\Theta(\tau_{14})(m/\tau_{14})^2  
   }{ \left(|\mathbb{S}^{d-1}|\Gamma(2-\frac{d}{2})\right)^2(r_1r_4)^{d-2}}\left\{\frac{2^{2 d-3}  m^{7-2 d} \tau _{14}^2}{|\mathbb{S}^{d-1}|  \Gamma \left(2-\frac{d}{2}\right)^2 \left(r_1
   r_4\right)^{d-2}}\int_{\tau_1>\tau_2>\tau_3>\tau_4}\frac{d\tau_2d\tau_3}{(\tau_{12}\tau_{34})^{4-d}\tau_{23}} \right.\hfill\\
   \times\int \frac{d r_2dr_3 }{(r_2 r_3)^{\frac{3d}{2}-4}}
  e^{-m
   \left(\frac{r_1^2+r_2^2}{\tau_{12}}+\frac{r_3^2+r_4^2}{\tau
   _{34}}+\frac{r_2^2+r_3^2}{\tau_{23}}\right)} \left[I_{\frac{d}{2}-1}\left(\frac{2m r_2 r_3}{\tau
   _{23}}\right)+O\left(r_1^{d-2},r_4^{d-2}\right)\right]\\
 +\frac{8 m^3 \tau_{14}^2}{|\mathbb{S}^{d-1}| }\sum_{l\geq 1}\frac{\text{deg}(d,l)C_l(\cos{\theta_{14}})}{(\Gamma (\frac{d}{2}+l))^2}\left(\frac{m^2 r_1r_4}{4}\right)^l \int_{\tau_1>\tau_2>\tau_3>\tau_4}\frac{d\tau_2d\tau_3}{(\tau_{12}\tau_{34})^{2+l}\tau_{23}} \\\left.
 \int d r_2dr_3 (r_2 r_3)^{2+l-\frac{d}{2}}  e^{-m
   \left(\frac{r_1^2+r_2^2}{\tau_{12}}+\frac{r_3^2+r_4^2}{\tau
   _{34}}+\frac{r_2^2+r_3^2}{\tau_{23}}\right)}\left[I_{\frac{d}{2}+l-1}\left(\frac{2m r_2 r_3}{\tau
   _{23}}\right)+O\left(r_1^{d-2},r_4^{d-2}\right)\right]\right\}\\
   =\frac{\Theta(\tau_{14})(m/\tau_{14})^2  
   }{ \left(|\mathbb{S}^{d-1}|\Gamma(2-\frac{d}{2})\right)^2(r_1r_4)^{d-2}}\left\{\frac{2^{2 d-5} m^{2-\frac{d}{2}}\tau _{14}^{\frac{3 d}{2}-2}}{|\mathbb{S}^{d-1}|  \Gamma\left(d/2\right) \left(r_1 r_4\right)^{d-2} }\right.\int_{\tau_1>\tau_2>\tau_3>\tau_4}\frac{d\tau_2 d\tau_3}{(\tau_{12}\tau_{34})^{\frac{d}{2}}}\hfill \\
\times e^{-m\left(\frac{r_1^2}{\tau_{12}}+\frac{r_4^2}{\tau_{34}}\right)}\left[\xi ^{d-2} \, _2F_1\left(d-2,d-2;\frac{d}{2};\xi \right)+O\left(r_1^{d-2},r_4^{d-2}\right)\right]\\
+\frac{2(m \tau _{14})^{\frac{d}{2}}}{|\mathbb{S}^{d-1}|} \sum_{l\geq 1}\frac{\text{deg}(d,l)C_l(\cos{\theta_{14}})(\Gamma(l+1))^2}{(\Gamma (\frac{d}{2}+l))^3}\left(\frac{m^2 r_1r_4}{4\tau_{14}}\right)^l\int_{\tau_1>\tau_2>\tau_3>\tau_4}\frac{d\tau_2 d\tau_3}{(\tau_{12}\tau_{34})^{\frac{d}{2}}}\\
\times e^{-m\left(\frac{r_1^2}{\tau_{12}}+\frac{r_4^2}{\tau_{34}}\right)}\left[\xi ^{\frac{d}{2}-1} \, _2F_1\left(\frac{d-2}{2},\frac{d-2}{2};\frac{d}{2}+l;\xi \right)+O\left(r_1^{d-2},r_4^{d-2}\right)\right]~,
\end{multlined}
\end{equation}
where $\xi\equiv \frac{\tau_{12}\tau_{34}}{\tau_{13}\tau_{24}}$ is the $SL(2,\mathbb{R})$ cross-ratio. The double integral $\int d\tau_2 d\tau_3$ in \eqref{appendix_d=4 defect anomalous dimension part 1} is evaluated as 
\begin{equation}
\label{appendix_d=4 defect anomalous dimension part 2}
\mathtoolsset{multlined-width=0.9\displaywidth}
\begin{multlined}
\int_{\tau_1>\tau_2>\tau_3>\tau_4}\frac{d\tau_2 d\tau_3}{(\tau_{12}\tau_{34})^{\frac{d}{2}}}e^{-m\left(\frac{r_1^2}{\tau_{12}}+\frac{r_4^2}{\tau_{34}}\right)}\xi ^{d-2} \, _2F_1\left(d-2,d-2;\frac{d}{2};\xi \right)\hfill \\
\hfill=-\frac{2\tau_{14}^{-\frac{d}{2}}}{3}\left[\ln{\frac{\tau_{14}}{mr_1r_4}}-\frac{1}{2}-\gamma_\text{E}+O\left(\epsilon,r_1^2,r_4^2\right)\right]\\
\int_{\tau_1>\tau_2>\tau_3>\tau_4}\frac{d\tau_2 d\tau_3}{(\tau_{12}\tau_{34})^{\frac{d}{2}}}e^{-m\left(\frac{r_1^2}{\tau_{12}}+\frac{r_4^2}{\tau_{34}}\right)}\xi ^{\frac{d}{2}-1} \, _2F_1\left(\frac{d-2}{2},\frac{d-2}{2};\frac{d}{2}+l;\xi \right)\hfill\\
=\frac{2\tau_{14}^{-\frac{d}{2}}}{l+1}\left[\ln {\frac{\tau_{14}}{m r_1 r_4}}-\gamma_\text{E}\right. \\
\hfill \left.+(l+1)\left((\log\frac{r_1}{r_4})^2+\frac{\pi^2}{12}-\psi ^{(1)}(l+1)+\frac{1}{(l+1)^2}\right)+O\left(\epsilon,r_1^2,r_4^2\right)\right]~,
\end{multlined}
\end{equation}
where $\psi^{(1)}(x)$ is the Polygamma function, and we have used dimensional regularization to the hypergeometric sum
\begin{equation}
\left.\, _3F_2\left(\frac{d}{2}-1,d-2,d-2;\frac{d}{2},\frac{d}{2};1\right)\right|_{d=4-\epsilon}=-\frac{2}{3}-\frac{5 \epsilon }{6}-\frac{5}{72} \left(18+\pi ^2\right) \epsilon ^2+O\left(\epsilon ^3\right)~.
\end{equation}
We remark that similar to \eqref{appendix_d=2 defect anomalous dimension part 2}, the integral \eqref{appendix_d=4 defect anomalous dimension part 2} is UV finite and implies no extra defect counterterms are needed. Plugging \eqref{appendix_d=4 defect anomalous dimension part 2} into \eqref{appendix_d=4 defect anomalous dimension part 1}, we obtain:
\begin{equation}
\label{appendix_d=4 defect anomalous dimension part 3}
\mathtoolsset{multlined-width=0.9\displaywidth}
\begin{multlined}
    \int \prod_{i=2}^3 d\tau_i d^dx_i [G_{m}(\tau_{12},x_{12})+D_m(\tau_{12},r_1,r_2)]^2\\
    \hfill \times G_{2m}(\tau_{23},x_{23})[G_{m}(\tau_{34},x_{34})+D_m(\tau_{34},r_3,r_4)]^2\\
    =\frac{\Theta(\tau_{14})(m/\tau_{14})^2  
   }{ \left(|\mathbb{S}^{d-1}|\Gamma(2-\frac{d}{2})\right)^2(r_1r_4)^{d-2}}\hfill \\
   \times \left[-\frac{4}{3}\left(\frac{mr_1r_4}{2\tau_{14}}\right)^{2-d} +4\sum_{l\geq 1}\frac{\text{deg}(d,l)C_l(\cos{\theta_{14}})}{(l+1)^3\Gamma (\frac{d}{2}+l)}\left(\frac{m^2 r_1r_4}{4\tau_{14}}\right)^l\right]\\
   \times\frac{m^2}{|\mathbb{S}^{d-1}|}\left[\ln {\frac{\tau_{14}}{m r_1 r_4}}+O\left((\ln {\frac{\tau_{14}}{m r_1 r_4}})^0,\epsilon\right)\right]~.
\end{multlined}
\end{equation}
From \eqref{appendix_defect tree level} and  \eqref{appendix_d=4 defect anomalous dimension part 3}, one can find the corrected defect scaling dimensions as in \eqref{eq_d=4 defect dimension spin 0} and \eqref{eq_d=4 defect dimension spin l}.

\bibliographystyle{unsrt}
\bibliography{ref}

\end{document}